\documentclass[reprint,amssymb,amsmath,aip,cha]{revtex4-1}
\usepackage{docs}
\usepackage{bm}
\usepackage{graphicx}

\usepackage[colorlinks = true,
            linkcolor = red,
            urlcolor  = red,
            citecolor = blue,
            anchorcolor = green]{hyperref}

\expandafter\ifx\csname package@font\endcsname\relax\else
 \expandafter\expandafter
 \expandafter\usepackage
 \expandafter\expandafter
 \expandafter{\csname package@font\endcsname}%
\fi
\begin{document}

\title{Coherent control of the refractive index using optical bistability}%

\author{H. Aswath Babu and Harshawardhan Wanare}%
%\author{H. Aswath Babu}
%\email{aswath@iitk.ac.in}
%
%\author{Harshawardhan Wanare}
\email{hwanare@iitk.ac.in}
\affiliation{Department of Physics, Indian Institute of Technology, Kanpur 208 016, India}%

\date{\today}%

\begin{abstract}
Refractive index and absorption experienced by a probe field propagating through a three-level atomic medium can be effectively manipulated by the bistable behavior of a control field. The probe field couples  the lower transition of the atom in ladder configuration and experiences normal or anomalous dispersion depending on the control field being in the upper or lower bistable state, respectively. We also obtain nonlinear dynamical instability in the form of periodic self-pulsing as the lower bistable branch becomes unstable, quite unlike earlier demonstrations of unstable regime in the upper branch. Consequently, the susceptibility experienced by the probe field varies periodically in time dictated by the control field self-pulsing.

\end{abstract}

\maketitle

%\tableofcontents

Interaction of light with matter in the linear regime is completely determined by the refractive index and absorption of the material. In the last two decades these two ubiquitous quantities have been engineered coherently  in a wide variety of ways resulting in  exciting phenomena such as slow-light~\cite{slight},  superluminal light~\cite{superluminal},  making a resonant absorbing medium 
transparent with electromagnetically induced transparency~\cite{eit}, achieving giant enhancements in the refractive index~\cite{scully}, coherently controllable metamaterials~\cite{ccontrol}, and many such developments that have transformed the practice of present day optics. Recently, coherent control of refractive index~\cite{chris2} and spatially varying refractive index accompanied by negligible absorption has also been proposed~\cite{chris}. The control of optical bistability (OB) using coherent interactions supplementing the bistable field has been in vogue for a few years now~\cite{harsha_ob, xiao-rapid}.  Recently, we have also predicted the existence  of negative hysteresis bistable response in three-level atom experiencing double feedback along  two adjacent transitions~\cite{ourpaper1}, and a host of  nonlinear dynamical behavior such as self-pulsing and chaos~\cite{ourpaper2}.  Nonlinear dynamics in driven OB systems involving three-level $\Lambda$ system in the electromagnetically induced transparency (EIT) regime has also been studied earlier~\cite{joshi2,joshi1}. Here, we present another mechanism that allows an effective manipulation of the refractive index using OB : varying from anomalous to normal as well as time-dependent periodic susceptibility (absorption and refractive index). Such manipulation is achieved through the cooperative effect at an adjacent transition exhibiting OB. However, our system does not rely on the EIT effect for the control mechanism. 

 Refractive index governs the frequency dependent phase delay experienced by the electromagnetic field in the medium, whereas the absorption coefficient its extinction. The Kramer-Kronig relations based on {\em causality} relate the spectral dependence of the  refractive index  and the associated  absorption, thus it is impossible to vary exclusively only the refractive index  or the absorption without affecting the other. The manipulation of  refractive index  and absorption through control field OB could be used to control the propagation of the probe field through the medium. We demonstrate that for the bistable field in the {\em ON}-state (large intra-cavity field associated with the upper branch of the S-shaped OB response~\cite{lugiato1984}) the probe field experiences normal dispersion accompanied with negligible absorption.
\begin{figure}[thb]
\includegraphics[width=8.5cm]{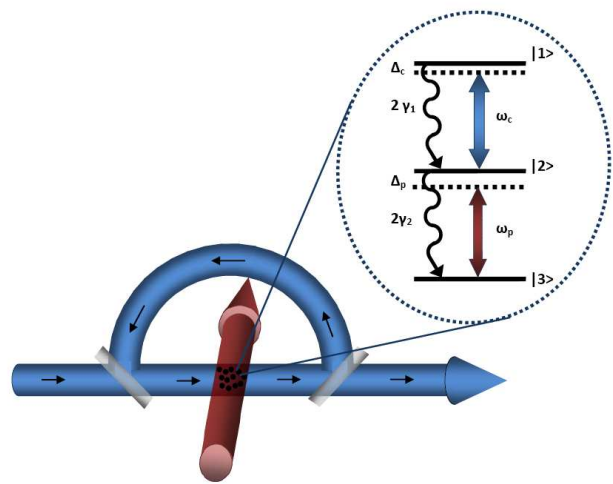}
\caption{ (Color online) The probe field (red) interacts with a collection of atoms whose susceptibility is dictated by the control field (blue) circulating in a cavity. (Inset) The three-level ladder ($\Xi$) system and the associated fields.}
\label{schatomfig}
\end{figure}
In contrast, the bistable field in the {\em OFF}-state (weak intra-cavity field associated with the lower cooperative branch) leads to anomalous dispersion and the associated absorption. In optical communication, apart  from being a switch~\cite{gibbs} OB can also provide an excellent handle to either delay (normal dispersion) or advance  (anomalous dispersion) of the  probe pulse. Furthermore, we propose a  scheme of realizing periodically varying susceptibility for the probe field by creating control field instability in the lower branch of OB. This instability occurs for a single mode field, at intensities much lower than  those required to saturate the atom and switch to the upper bistable state of OB.  The creation of  instability, assisted by atomic coherence leads to nonlinear dynamical effects, such as self-pulsing whose period and amplitude can be tuned to obtain a range of time-scales for the susceptibility. 

 Earlier studies of instability  in OB systems range from the Ikeda instability~\cite{ikeda} in two-level OB systems~\cite{lambrecht} to  three-level driven OB systems \cite{xiao-rapid,xiao2}. We believe that the regime of instability reported here is intrinsically different from these earlier reports. The self-pulsing in the upper branch reported earlier arises due to a competition between two time scales, one associated with the slow population transfer through optical pumping and another involving fast variation of the optical nonlinearities as explained in the Ref.~\onlinecite{lambrecht,xiao-rapid}. In our system, the self-pulsing occurs at low input intensities, sometimes even lower than intensities associated with the bistable thresholds. Moreover, it does not necessitate finite detuning of either the bistable field or the cavity detuning~\cite{carmichael}, and occurs without involving coupling to multiple modes of the cavity.
\begin{figure}[htb]
%\centering
\hspace*{-1cm}
\includegraphics[width=10cm,height=6.5cm]{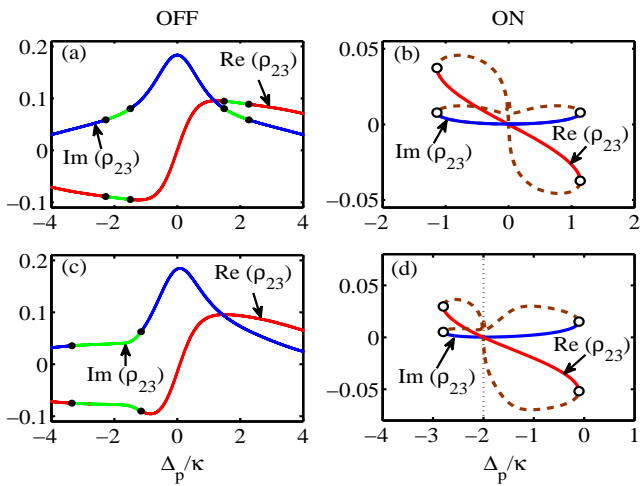}
\caption{(Color online) (a) The refractive index (red) and the absorption (blue) experienced by the probe field corresponding to the lower branch of OB and similarly (b) shows the probe susceptibility for the upper branch of OB. (c) and (d) indicate the probe susceptibility for the detuned control field $\Delta_c/\kappa=2$. The stable regions are separated from the unstable by the Hopf (H) and Limit points (LP) indicated with dots and open circles, respectively. The control field input $|y_{c}|=3.0$, and the other parameters are $C=200,|G_{p}|/\kappa=0.2$, $\gamma_1/\kappa=0.01$,  $\gamma_2/\kappa=1$, and $ \delta_c=0$.}
\label{norm_anom}
\end{figure}

 We consider an unidirectional optical ring cavity of total length $\cal L$ and resonant cavity frequency $\omega_o$, having the active medium confined within length $L$ and with mirrors of transitivity $T$, as shown in Fig.~\ref{schatomfig}. The active medium consists of a closed atomic system in the $\Xi$ (ladder) configuration, and couples to a weak probe (having frequency $\omega_p$) and strong control field (having frequency $\omega_c$)  at lower and upper transitions, respectively. Only the control field circulates in the cavity and exhibits cooperative phenomenon as it experience sufficient feedback provided by the cavity. The normalized cavity detuning with respect to control field is defined as $\delta_c=(\omega_o-\omega_c){\cal L}/c$. For simplicity, we consider the OB phenomenon in the mean field limit ({\em i.e.} $\alpha L \rightarrow 0$, $T \rightarrow 0 $, such that $C=\alpha L/2 T$ is finite)~\cite{lugiato1984}. This ensures that the control field experiences weak absorption ($\alpha$: absorption coefficient along the upper transition) in a single traversal through the cavity and it undergoes many such round trips in the cavity leading to substantial interaction, moreover, the field distribution remains spatially uniform throughout the active medium.  Here, the probe field is weak, and  thus  couples to the atom linearly, whereas the atom-control field coupling OB is considered to all orders. However, we have not undertaken any such perturbative truncation, and the results presented here were obtained from numerical simulations undertaken along the lines described in detail in Ref.~\onlinecite{ourpaper2}. 
 
 We consider only the simplest  configuration with regard to the atom-cavity coupling that captures the essential physics, where  the atoms are homogeneously broadened exhibiting radiative decay, moreover the atoms couple to a single mode of the cavity. We do not consider inhomogeneously broadened media here~\cite{tdb}. The atom-field interaction is described by the density-matrix equation Eq.~\ref{density_mat} in the semi-classical regime, and the field dynamics are described by Eq.~\ref{boundary} obtained in the mean field limit under slowly varying envelope approximation, which captures the cavity feedback applied to the control field in terms of the cooperative parameter ($C$).
\begin{eqnarray}
\label{density_mat}
\frac{\partial\rho}{\partial t} &=&-\frac{i}{\hbar} \left[{\hat H},\rho \right]+\hat{ \cal{ L}}
\rho,\\
\frac{\partial x_c}{\partial  t}& = & \kappa \left[-\left(1+i \frac{\delta_c}{T}\right)x_c + y_{c} + 2 i C \rho_{12}\right].
\label{boundary}
\end{eqnarray}
The atom-field coupling and the atom-field detunings are contained in the total Hamiltonian $\hat H$ $= \hbar[(\Delta_c + \Delta_p) |1\rangle \langle 1| + \Delta_p |2\rangle \langle 2| -(G_c |1\rangle \langle 2| + G_p |2\rangle \langle 3| + \rm{H.c.})]$  in the dipole approximation after undertaking the rotating wave approximation. The probe field coupling is given by the Rabi frequency $G_{p}  = \vec d_{23} \cdot \vec E_{p}/\hbar$ and the detuning is $\Delta_p = \omega_{23}-\omega_{p}$, where $\vec E_{p}$ is the probe field amplitude. Similarly, we define $G_c$ and $\Delta_c$ associated with the intra-cavity control field. The incoherent processes such as spontaneous emission decays ($2\gamma_i$) from the state $|i\rangle$, as shown in Fig.~\ref{schatomfig}, are contained in the Liouville operator ($\hat{ \cal{ L}}$). The normalized cavity input and output strength of the control field are defined as $y_{c}= {\vec d_{12} \cdot \vec E_{c}^{in}}{/}{(\hbar\kappa \sqrt{T})}$ and $x_c = {\vec d_{12}\cdot \vec E_{c}^{out}}{/}{(\hbar \kappa \sqrt{T})}$, respectively.
%$y_{c}$ and the output field $x_c$ are given as
%\begin{eqnarray}
%$y_{c}= {\vec d_{12} \cdot \vec E_{c}^{in}}{/}{\left(\hbar\kappa \sqrt{T}\right)}$,
%$x_c = {\vec d_{12}\cdot \vec E_{c}^{out}}{/}{\left(\hbar \kappa \sqrt{T}\right)}$,
%\end{eqnarray}
Here $\vec d_{12}$ is the dipole moment associated with the
$|1\rangle \leftrightarrow |2\rangle$ transition and $ \vec E_{c}^{in(out)}$ is the amplitude of the input (output) control field.  All the frequency units are normalized with respect to the cavity decay $\kappa$, unless specified otherwise. 

 The probe susceptibility can be obtained under two circumstances depending on the specific state of the bistable control field. These bistable states correspond to the cooperative (lower) branch and the one-atom (upper) branch of the S shaped OB response. The refractive index and absorption is directly proportional to the real and imaginary parts of $\rho_{23}$, respectively, and their dependence on the probe detuning $\Delta_p$ is presented in Fig.~\ref{norm_anom} ({\em OFF/ON}-state in the left/right panel). Note that the input control field strength and its detuning is held constant as the probe frequency is varied. For the system in the lower OB branch the probe field experiences anomalous dispersion accompanied by an absorption peak at $\Delta_p/\kappa = 0$ (Fig.~\ref{norm_anom} (a)), whereas for the system in the upper branch the probe field experiences normal dispersion accompanied with negligible absorption (Fig.~\ref{norm_anom} (b)). For finite control field detuning $\Delta_c/\kappa=2$, the probe response is given in Fig.~\ref{norm_anom} (c, d).
 \begin{figure}[thb]
\centering
\includegraphics[width=0.50\textwidth]{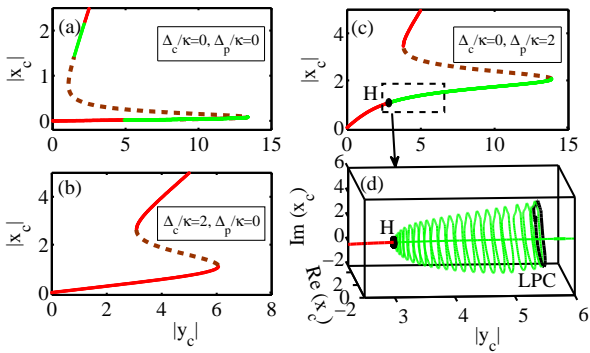}
\caption{ (Color online) Optical bistable response of the control field as $y_c$ is varied for various detunings. (a) $\Delta_c/\kappa=0$, $\Delta_p/\kappa=0$, (b) $\Delta_c/\kappa=2$, $\Delta_p/\kappa=0$, (c) $\Delta_c/\kappa=0$, $\Delta_p/\kappa=2$. The nonlinear dynamical regime is shown in green. (d) The limit cycle continuation from the Hopf point ($H$) in (c) leads to periodic self-pulsing. The Limit point of cycle ($LPC$) is indicated in black and  self-pulsing does not occur beyond it. The involved parameters are same as in Fig.~\ref{norm_anom}}
\label{bis_norm_anom}
\end{figure} 
 
 Apart from obtaining the steady state response of $\rho_{23}$, we have undertaken the linear stability analysis~\cite{kuznetsov} where the eigenvalues of the Jacobian matrix of the system are examined to identify stable and unstable steady states indicated in red/blue and green/brown curves throughout the manuscript. The physically inaccessible unstable states having eigenvalues with positive real and zero imaginary parts are shown with brown curves. The regime of nonlinear dynamical instability are indicated in green color curves. The steady state curve of the real part of $\rho_{23}$ corresponding to the bistable region of the control field is in the form of a closed loop (Fig.~\ref{norm_anom} (b)), where the dashed brown curve  region is inaccessible corresponding to the usual unstable solution of the S-shaped OB curve, and the system switches to available {\em ON} or {\em OFF}-state. We would like to point out that corresponding to each point in Fig.~\ref{norm_anom} one can obtain the associated S-shaped OB response (Fig.~\ref{bis_norm_anom}) as the control field input strength is varied.
 
\begin{figure}[thb]
%\centering
\hspace*{-0.7cm}
\includegraphics[width=9.8cm,height=6.3cm]{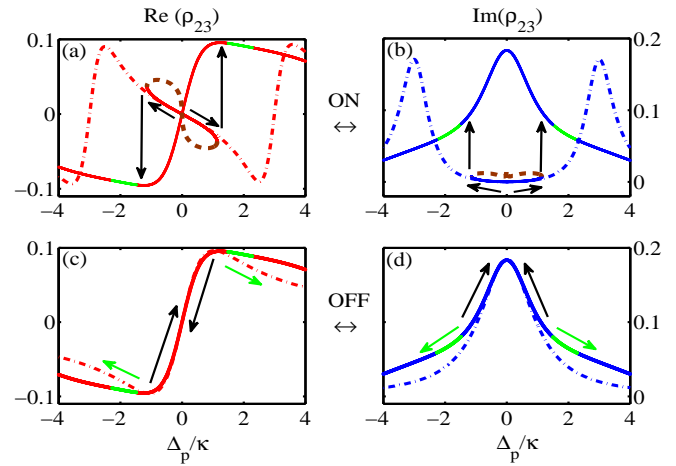}
\caption{ (Color online) The switching of refractive index/ absorption as the probe detuning is varied. (a) \& (b) corresponds to normal dispersion and indicates the switching from the {\em ON}-state to {\em OFF} state. (c) \& (d) corresponds to the {\em OFF}-state exhibiting anomalous dispersion (black arrows). The green arrows indicate the probe frequency variation leading to periodic self-pulsing: The dash-dotted lines correspond to response without feedback. The parameters are same as Fig.~\ref{norm_anom}}
\label{norm_anom_eit}
\end{figure}

 In order to understand the effects of the cavity feedback, we have compared the probe dispersion and absorption with the conventional response arising without feedback. In Fig.~\ref{norm_anom_eit}(a) and (b) the dash-dotted curves indicate the probe response without feedback for $|y_c|=|G_c|/\kappa=3$, which is the intra-cavity value for OB in the upper branch, where an EIT like profile is observed. The arrows indicate how the system switches from the top branch to the lower branch as the probe detuning $\Delta_p$ is varied. In order to transform the associated dispersion from the normal to anomalous regime one need to access the lower branch, and which can be achieved by adiabatically changing the probe detuning along the upper branch beyond the turning points (Fig.~\ref{norm_anom_eit}(a)), and the system would jump to the lower OB state exhibiting anomalous dispersion (Fig.~\ref{norm_anom_eit}(c)).

 The above response is consistent with the analytical results obtained in the perturbative limit of a weak probe field.  The  change in the slope of the dispersion is clear from expression of the density matrix element
 \begin{equation}
\rho_{23}={i G_p}{\bigg /} {\left[ \gamma_2+i \Delta_p+ \frac{|G_c|^2}{\gamma_1+i
(\Delta_c+\Delta_p)}\right]}.
\label{r23}
\end{equation}
The control field within the cavity dictates the probe response. For absorptive OB case ($\Delta_c/\kappa = 0$), when the system is in the lower branch, the intra-cavity field would be weak due to the large collective absorption of the atoms and that results in a conventional Lorentzian response for the probe field, as the $|G_c|^2$ term in Eq.~\ref{r23} is negligible.  However, when the system is in the upper branch the term $|G_c|^2$ is dominant and the imaginary part of $\rho_{23}$ that becomes negligible. In dispersive OB case ($\Delta_c/\kappa = 2$), it is apparent that the term $|G_c|^2$ becomes even more significant at the two-photon resonance, {\em i.e.} $\Delta_c+\Delta_p=0$. This is clear from Fig.~\ref{norm_anom} (c) \& (d), wherein the upper branch response occurs only at the two-photon resonance, unlike for the system in the lower branch which continues to peak at the probe resonance $\Delta_p \approx 0$. The probe continues to experience anomalous dispersion at $\Delta_p \approx 0$ and normal dispersion at $\Delta_p \approx -\Delta_c$. The change in sign of the dispersion profile is clearly seen in the derivative of $\rho_{23}$ with respect to the detuning $\Delta_p$, in the vicinity of $\Delta_p \approx 0$, and at $\Delta_p \approx -\Delta_c$ for appropriate intra-cavity control field strength,
\begin{equation}
\frac{d \rho_{23}}{d \Delta_p} = \frac{G_p
\left[1-\frac{|G_c|^2}{\gamma_1+ i 
(\Delta_c+\Delta_p)^2}\right]}{\left[\gamma_2+i
\Delta_p+\frac{|G_c|^2}{\gamma_1+i (\Delta_c+\Delta_p)}\right]^2}.
\end{equation}
However, the complete bistable response arises from  cavity feedback which is not included in the above analytical expressions. It should also be noted that the population is largely confined to the ground state $|3\rangle$, with only about $10^{-1} \sim 10^{-2}$ in the excited states when the atom is in the lower branch of OB. For the atom in the upper branch, the steady state populations of states $|1\rangle$ and $|2\rangle$ are about $10^{-4}$ and $10^{-6}$, respectively. Moreover, in the top branch the ratio of steady state populations is related to the ratio of decay rates $\rho_{11}/\rho_{22} \propto \gamma_1/\gamma_2$. Such a three-level $\Xi$ system could be realized in $Rb^{85}$ vapor along the $5S_{1/2}\rightarrow 5P_{3/2}\rightarrow 5D_{5/2}$ transitions with a number density of $\sim 10^{17}$ atoms/m$^3$ and spontaneous emission decay rate $\sim 10^7 Hz$, in a ring cavity with $T \sim 10^{-2}$, and the resulting cooperative parameter $C\sim 1000$. The input power levels for the bistable field $\sim 20 mW$ across a spot size of $100 \mu m$ would be sufficient for switching to the upper branch. The probe susceptibility and the associated group velocity are given as:
\begin{eqnarray}
\chi = \frac{3 N \lambda_{p}^3 }{8 \pi^2}  \left(\frac{\gamma_2\rho_{23}}{G_p} \right),\hspace{0.1cm} v_g =\frac{c}{n_g},
\label{group_velocity}
\end{eqnarray}  
where the group index $n_g=n + \omega_p ({\partial n}/{\partial \omega_p})$. The relation between the refractive index ($n$) and the susceptibility ($\chi$) is given by $n^2=n_{bg}^2+\chi$, where $n_{bg}$ is the background refractive index. 
%Using 
%\begin{eqnarray}
%\frac{\partial n}{\partial \omega_p}=-\frac{1}{2}\frac{\partial \chi}{\partial \Delta_{p}} &=&-\frac{3 N \lambda^3_{p} \gamma_2 }{16 \pi^2 G_p} \left(\frac{\partial \rho_{23}}{\partial \Delta_{p}}\right),
%\end{eqnarray} 
Taking an atomic number density $N$ $\sim$ $6 \times 10^{17}$ atoms/m$^3$, spontaneous decay rate $\gamma_2$ $\sim$ $2 \times 10^7$Hz, the wavelength of the probe laser $\lambda_p$ $\sim$ $780$ nm, and  $\partial \rho_{23}/ {\partial \Delta_p}$ obtained from Fig.~\ref{norm_anom} (a) \& (b), result in a group index for both the anomalous and normal dispersion conditions as $\sim$ $-15 \times 10^4$ and $4.3 \times 10^4$, respectively. 

    We now focus our attention to the effects of feedback of the control field that results in nonlinear dynamical self-pulsing. The probe response (solid) with the control field OB feedback digresses from the non-cavity case (dash-dot) for $|\Delta_p|/\kappa$ $>$ $\gamma_2/\kappa$ as seen in Fig.~\ref{norm_anom_eit}. We would like to emphasize that this difference arises from the feedback and lies at the heart of the periodic self-pulsing (green color region) that the system exhibits in the lower branch. 

 The atomic system is chosen such that the uppermost state $|1\rangle$ is metastable and thus has a longer lifetime in comparison to the intermediate state $|2\rangle$. The stability of the lower bistable branch depends on the population injection by the probe field into the bistable transition. Such processes leads to the creation of instability in the lower bistable branch for appropriate probe detuning (see Fig.~\ref{bis_norm_anom}(a) and~\ref{bis_norm_anom}(c)). This unstable regime is associated with oscillatory dynamics such as stable periodic self-pulsing (see Fig.~\ref{bis_norm_anom}(d)). In our parameter regime the self-pulsing does not necessarily occur between the lower and upper branches of the bistable response~\cite{carmichael}. We believe that the physical mechanism involved in our system is intrinsically different from earlier reports of instability in the upper branch where optical pumping is invariably involved which relies on slow processes such as spontaneous emission~\cite{lambrecht}. In contrast, our system involves a competition between the population decay from the state $|2\rangle$ to the ground state $|3\rangle$ and nonlinear atom-field interaction along the transition $|1\rangle \leftrightarrow |2\rangle$ of the control field experiencing sufficient feedback. Atomic coherence is essential in obtaining the oscillatory  behavior, and inclusion of larger decoherence of  the atomic polarizations ($\rho_{12}$ or $\rho_{23}$) leads to a reduction of the unstable range and its eventual disappearance. It should be noted that obtaining bistable  solution and the associated switching  between the states is not essential for obtaining nonlinear dynamical self-pulsing in the lower branch, however, feedback (cooperative effect) is mandatory. Furthermore, self-pulsing could occur at the very onset of the lower cooperative branch, thus requiring much lower input intensities in contrast to the conventional single mode self-pulsing that occurs in the upper branch~\cite{lugiato_self}. The concerns over power requirement can further be relaxed by implementing this scheme using photonic crystal based micro-cavities which provide significant field enhancements~\cite{pcrystal} accompanied by short cavity round trip times. 
 
\begin{figure}[thb]
\centering
\includegraphics[width=0.48\textwidth]{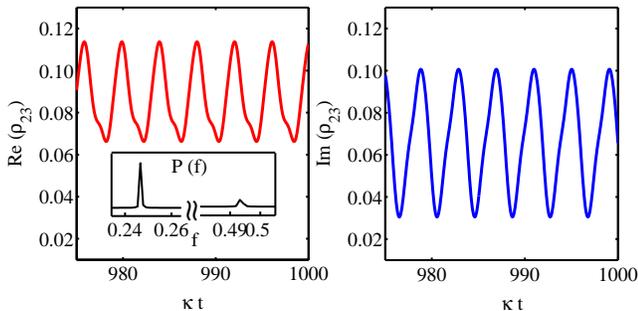}
\caption{(Color online) The time-periodic self-pulsing of the real and imaginary parts of $\rho_{23}$ and the power spectral density (Inset) are shown for $\Delta_p/\kappa = 2$, $|y_{c}| = 3$, $G_c/\kappa=0.2$ and all the other parameters are same as in Fig.~\ref{norm_anom}.}
\label{ladder_time}
\end{figure}

 The self-pulsing domain associated with the unstable solutions are separated from the stable solutions by either a Hopf (super-critical) point or a Limit point, as indicated in Fig.~\ref{norm_anom} with green/brown curve. The limit cycles that originate from the Hopf points are robust and stable. The Floquet analysis indicates that the Floquet multipliers are bound by unity, within this range of detunings, leading to stable limit cycles \cite{strogatz}. The stable  periodic oscillations occur for both the absorption as well as the refractive index (see Fig.~\ref{ladder_time}). The observed modulation in the refractive index and the absorption is of the order of $3 \% $ and $7 \% $, respectively. These oscillations are not necessarily sinusoidal, and one can obtain a variety of multiply peaked periodic oscillations with varied frequency content. The self-pulsing time scale is  governed by the cavity decay time $\kappa^{-1}={\cal L}/cT$. In our simulations the atomic states are long lived in comparison to cavity decay times, and in order to obtain this phenomenon, which can be chosen to have a  variety of values  satisfying the condition $\gamma_1 \ll \gamma_2 \le  \kappa$. In particular, for a micro-cavity set-up we consider $\gamma_1/ \kappa \sim 10^{-7}$,  $\gamma_2/\kappa \sim 10^{-5}$, $L \sim 10^{-6}$ m, $T \sim 10^{-2}$, $C \sim 20$ and obtain similar self-pulsing, wherein $\kappa \sim 10^{12}$ Hz and thus terahertz periodic self-pulsing seems possible.

In conclusion, we have demonstrated a coherent way to engineer refractive index (normal to anomalous) and absorption of a probe field using OB. Such OB in multi-level system can be used as optical delay lines for control of pulse propagation in integrated optical devices. Furthermore, we also demonstrate a mechanism to obtain time-dependent oscillatory susceptibility, with constant input control field, operating in the lower cooperative branch of OB. A distinct possibility of exploiting this scheme to obtain refractive index modulation in the terahertz regime needs to be tested, as the conventional slow dynamics associated with optical pumping is no longer a limiting factor in our scheme.


\begin{thebibliography}{99}
\bibitem{tdb} In gas media, the two photon Doppler broadening can be minimized by choosing counter-propagating beams for the atoms in the ladder ($\Xi$) configuration.
\end{thebibliography}
\end{document}